\documentclass[twocolumn,showpacs,preprintnumbers,amsmath,amssymb,floatfix,prd]{revtex4}
\usepackage{epsfig}
\usepackage{dcolumn}
\begin{document}
\title{Uncertainties in Pion and Kaon Fragmentation Functions}
\author{Manuel Epele}\email{manuepele@gmail.com}
\affiliation{ Instituto de F\'{\i}sica La Plata, CONICET - UNLP,
Departamento de F\'{\i}sica,  Facultad de Ciencias Exactas, Universidad de
La Plata, C.C. 69, La Plata, Argentina}
\author{Romina Llubaroff}\email{rll@df.uba.ar}
\affiliation{Departamento de F\'{\i}sica and IFIBA,  Facultad de Ciencias Exactas y Naturales, Universidad de Buenos Aires, Ciudad Universitaria, Pabell\'on\ 1 (1428) Buenos Aires, Argentina}
\author{Rodolfo Sassot}\email{sassot@df.uba.ar}
\affiliation{Departamento de F\'{\i}sica and IFIBA,  Facultad de Ciencias Exactas y Naturales, Universidad de Buenos Aires, Ciudad Universitaria, Pabell\'on\ 1 (1428) Buenos Aires, Argentina}
\author{Marco Stratmann}\email{marco@bnl.gov}
\affiliation{Physics Department, Brookhaven National Laboratory, Upton, NY~11973, USA}

\begin{abstract}
We present a detailed assessment of uncertainties in parton-to-pion and parton-to-kaon 
fragmentation functions obtained in recent global QCD analyses of single-inclusive hadron production data  
at next-to-leading order accuracy. 
We use the robust Lagrange multiplier approach for determining uncertainties to validate
the applicability of the simpler but approximate Hessian method.
Extensive comparisons of the results obtained within both methods are presented 
for the individual parton-to-pion and kaon fragmentation functions.
We provide Hessian eigenvector sets of pion and kaon fragmentation functions
that allow one to easily propagate their uncertainties to any observable. 
Various applications of these sets are presented for pion and kaon production
in electron-positron annihilation, lepton-nucleon scattering, and proton-proton collisions.
\end{abstract}


\maketitle

\section{Introduction and Motivation}
%
Fragmentation functions (FFs) contain vital non-perturbative information required to describe
a great variety of hard scattering processes with a given identified hadron in the final-state 
within the framework of perturbative Quantum Chromodynamics (pQCD).
The enhanced sensitivity of such less inclusive probes to crucial features of the 
partonic structure of hadrons, including its flavor content, dependence on spin 
\cite{deFlorian:2008mr,Leader:2010rb}, 
or possible nuclear modifications \cite{Sassot:2009sh,deFlorian:2011fp}, 
combined with ever growing experimental precision,
opens up novel opportunities for phenomenological QCD studies with an accuracy 
hitherto reserved to fully inclusive measurements.
In this context, the availability of reliable sets of parton-to-hadron FFs, 
as well as accurate estimates of their uncertainties, is of the utmost relevance. 

The extraction of FFs from data, originally restricted to 
single-inclusive electron-positron annihilation (SIA) \cite{ref:kretzer,ref:kkp,ref:akk,Hirai:2007cx}, 
has evolved in recent years into truly global QCD analyses encompassing 
data obtained in hadron-hadron collisions \cite{deFlorian:2007aj,de Florian:2007hc,Albino:2008fy}
as well as hadron multiplicities in semi-inclusive deep-inelastic scattering (SIDIS) \cite{deFlorian:2007aj}. 
Only the combined analysis of the large body of existing data taken in different processes and for different hadron species
allows for a consistent determination of all aspects of hadronization such as fully charge and flavor
separated parton-to-hadron FFs. 
A global fit also constitutes an explicit check of the assumed underlying factorizability and universality 
of FFs in hard processes, which is the foundation of the predictive power of pQCD.

As is the case with parton distribution functions (PDFs) \cite{ref:pdf}, the assessment 
of uncertainties for FFs obtained in global QCD analyses is far from 
being straightforward. 
Such studies incorporate data from many different experiments with diverse 
characteristics and errors. The data are in turn confronted with theoretical estimates 
whose inherent uncertainties are notoriously difficult to quantify, because, 
in addition to the truncation of the perturbative expansion at a given order of pQCD, 
there are also unavoidable approximations and assumptions involved. 
The latter comprise the choice of the functional form used to parametrize FFs \cite{Albino:2008aa} 
(or PDFs \cite{Alekhin:2011sk})
at some low input scale $Q_0\simeq 1\,\mathrm{GeV}$, the value of the strong coupling
at some reference scale, the selection of and cuts applied to the data sets used in the fit, 
or the treatment of heavy flavors in the scale evolution and the calculation of observables.

In recent years, significant progress has been made in putting forward,
exploring, validating, and comparing different strategies to estimate uncertainties 
in global QCD analyses of PDFs.
Among the various approaches, the robust Lagrange Multiplier (LM) technique \cite{Stump:2001gu}
directly relates the variation of the parameters determined in the fit or, more generally, 
of any observable computed with them, to the variation of the $\chi^2$ function that quantifies 
the goodness of the fit to data. 
While this avoids any approximations or assumptions about the dependence of the 
$\chi^2$ hypersurface on the parameters used to describe the PDFs or how to 
propagate uncertainties to a given observable, it requires an extensive amount
of $\chi^2$ minimizations involving all data sets included in the fit.
The computationally less demanding ``Improved (iterative) Hessian'' (IH) approach \cite{Pumplin:2001ct}
assumes a quadratic behavior of the $\chi^2$ hypersurface on the parameter displacements and represents the 
$\chi^2$ increment from its minimum value in terms of combinations of the fit parameters that maximize the variation. 
Within this eigenvector representation of the Hessian matrix, 
and assuming a linear propagation of errors, the uncertainty of any observable can be 
straightforwardly estimated from a set of pre-calculated fits 
corresponding to fixed displacements along the eigenvector directions. 
Another approach is based on analyzing a large amount of replicas of the original
data sets with neural networks \cite{Ball:2010de}. While this has the drawback of defining the central
fit only as the statistical average of ${\cal{O}}(100)$ PDF fits, it is largely free
of the bias from assuming a certain functional form for the PDFs at scale $Q_0$.

Corresponding studies of uncertainties in parton-to-hadron FFs are 
scarce despite the large body of available data.
An analysis based solely on SIA data, which do not allow for a full charge and
flavor separation of FFs, was presented in Ref.~\cite{Hirai:2007cx} using the Hessian 
method but without validating its applicability.
The so far most comprehensive analysis based on data from SIA, SIDIS, and
hadronic collisions was presented in Refs.~\cite{deFlorian:2007aj,de Florian:2007hc}, 
henceforth referred to as DSS FFs. However, uncertainties of the extracted FFs 
were only assessed qualitatively for certain truncated moments within the
LM approach. 

In the following, we will extend the DSS analysis by performing a detailed 
assessment of the uncertainties in pion and kaon FFs based on the IH method. 
We closely follow the DSS framework \cite{deFlorian:2007aj} 
and adopt the same functional form
to parametrize the FFs at the initial scale $Q_0=1\,\mathrm{GeV}$ and
the same selection of data sets.
In general, we find good agreement with the results obtained with the
robust LM technique but notice that for the same nominal tolerance criterion,
i.e., increase in $\chi^2$, the IH method typically leads to somewhat smaller
uncertainty estimates. We devise a recipe to account for these small differences
and provide sets of Hessian eigenvectors FFs to facilitate the propagation of
uncertainties to arbitrary observables.
We believe that such an analysis is particularly timely and useful in view of
the wealth of upcoming or new precise data on identified hadron yields in
SIA from $B$ factories \cite{ref:belleprel}, SIDIS \cite{ref:compassmult},
and hadron-hadron collisions both at RHIC \cite{Agakishiev:2011dc} and 
the LHC \cite{Abelev:2012cn}.
The results of our analysis will help to quantify the impact of these data sets,
identify possible tensions with the DSS analysis, and will serve as the baseline result 
for an anticipated update of the DSS sets of pion and kaon FFs.

The remainder of the paper is organized as follows: in the next Section we briefly 
recall the main aspects of the DSS analysis and the IH method, study in detail the uncertainties
of parton-to-pion FFs in IH approach, and compare to the results 
obtained with the LM technique.
Hessian eigenvector sets of pion FF are provided and applied to calculations of
pion yields in SIA, SIDIS, and hadron-hadron collisions.
Section III is devoted to a similar study for kaon FFs. We briefly summarize the
main results in Sec.~IV.

\section{Pion Fragmentation Functions}
\subsection{Preliminaries \label{sec:prel}}
%
Since both the framework and methodology for the extraction of FFs in global QCD 
analyses at next-to-leading order (NLO) accuracy as well as the implementation of the IH and LM techniques 
have already been explained in quite some depth in the literature \cite{deFlorian:2007aj,Stump:2001gu,Pumplin:2001ct}, 
we will only briefly recall the main concepts and results relevant for our studies.

As was mentioned in the Introduction, we choose the DSS NLO analysis of pion and kaon FF
\cite{deFlorian:2007aj} and their error estimates based on the LM method 
as our baseline fit. 
We adopt the same selection of data sets used in the DSS fit and the
same flexible functional form
\begin{equation}
\label{eq:ff-input}
D_i^H(z,Q_0) =
\frac{N_i z^{\alpha_i}(1-z)^{\beta_i} [1+\gamma_i (1-z)^{\delta_i}] }
{B[2+\alpha_i,\beta_i+1]+\gamma_i B[2+\alpha_i,\beta_i+\delta_i+1]},
\end{equation}
to parametrize the hadronization of a parton $i$ into a hadron $H$ at 
the initial scale $Q_0=1\,\mathrm{GeV}$. $z$ denotes the fraction
of the parton's momentum taken by the observed hadron, and
$B[a,b]$ represents the Euler Beta-function. The $N_i$ in (\ref{eq:ff-input})
are normalized such to represent the contribution of $D_i^H$ to the 
momentum sum rule 
\begin{equation}
\label{eq:sumrule}
\sum_H \int_0^1 dz z D_i^H(z,Q^2) = 1\,.
\end{equation}
Since presently available data do not constrain all free parameters in (\ref{eq:ff-input})
for each parton $i$ equally well, certain relations upon the individual FFs had to be 
imposed in the DSS analysis \cite{deFlorian:2007aj} without jeopardizing the quality of the fit.
Apart from assuming isospin symmetry for the unfavored sea quark FFs, i.e.,
$D_{\bar{u}}^{\pi^+}=D_{d}^{\pi^+}$, the total $u$-quark $D_{d+\bar{d}}^{\pi^+}$ and
$d$-quark $D_{u+\bar{u}}^{\pi^+}$ FFs are only allowed to differ in normalization $N$, and
the strange quark FFs $D_s^{\pi^+}=D_{\bar{s}}^{\pi^+}$ are related to the sea
quark FFs through another normalization factor $N^{\prime}$. The corresponding FFs
for negatively charged pions are obtained by charge conjugation invariance and
those for neutral pions by assuming $D_i^{\pi^0}= [D_i^{\pi^+}+D_i^{\pi^-}]/2$.

The remaining 23 free parameters $\{a_i\}$ describing the DSS FFs 
for quarks and gluons into positively charged pions, $D_i^{\pi^{+}}$,
are then determined by a standard $\chi^2$ minimization.
Since the full error correlation matrices are not available for
most of the data sets used in the fit, statistical and systematical errors 
are usually added in quadrature \cite{ref:kretzer,ref:kkp,ref:akk,Hirai:2007cx,deFlorian:2007aj}.
As in the DSS analysis, we allow each data set to float within the quoted experimental normalization
uncertainty by introducing a set of 7 extra parameters to the fit.

It suffices to say that we fully reproduce the set of optimum parameters $\{a_i^0\}$ of the
DSS analysis \cite{deFlorian:2007aj}, corresponding to the minimum in the $\chi^2$ profile.
To explore their uncertainties with the IH approach \cite{Pumplin:2001ct} we express the
Hessian matrix
\begin{equation}
\label{eq:hij}
H_{ij} \equiv \frac{1}{2} \frac{\partial^2 \chi^2}{\partial y_i \partial y_j}
\Bigg|_{0} \;\;,
\end{equation}
where the derivates are taken at the minimum, 
in terms of its $N_{\mathrm{par}}$ eigenvectors $v_i^{(k)}$. 
The displacements $y_i \equiv  a_i - a_i^0$ in Eq.~(\ref{eq:hij}) 
and the increase in $\chi^2$ 
\begin{equation}
\label{eq:chi2hessian}
\Delta \chi^2 =  \chi^2(\{a_i\}) - \chi^2_0(\{a_i^0\}) = 
\sum_{ij} H_{ij} y_i y_j
\end{equation}
are then replaced by a new set of 
parameters $\{z_i\}$ defined by \cite{Pumplin:2001ct}
\begin{equation}
\label{eq:zi-def}
y_i \equiv \sum_j v_i^{(j)} s_j z_j\,.
\end{equation} 
The factors $s_j\propto \sqrt{1/\varepsilon_j}$ are used to
rescale the $\{z_i\}$ such that the distance from the 
$\chi^2$ minimum is simply given by
\begin{equation}
\label{eq:deltachi2z}
\Delta \chi^2=  \sum_i z_i^2\,.
\end{equation}
The $\chi^2$ function changes rapidly for directions corresponding to large
eigenvalues $\varepsilon_k$ of the Hessian matrix while small eigenvalues
belong to directions where the fit parameters are only weakly constrained.

What makes the eigenvector representation $\{z_i\}$ particularly useful and 
convenient is the possibility to construct 2$N_{\mathrm{par}}$ 
basis sets $S_k^{\pm}$ of FFs which greatly facilitate the propagation of their
uncertainties to arbitrary observables ${\cal{O}}$.
An estimate of the error $\Delta{\cal{O}}$ away from its 
best fit estimate ${\cal{O}}(S^0)$ is obtained by computing \cite{Pumplin:2001ct}
\begin{equation}
\label{eq:obserror-hessian}
\Delta{\cal{O}} = \frac{1}{2} \left[ \sum_{k=1}^{N_{\mathrm{par}}}
[{\cal{O}}(S^+_k) - {\cal{O}}(S^-_k)]^2 \right]^{1/2}\;.
\end{equation}
The eigenvector sets $S_k^{\pm}$ are defined by choosing the amount
$T=\sqrt{\Delta \chi^2}$ still tolerated for an acceptable global fit
and correspond to positive and negative displacements by $T$
along each of the eigenvector directions 
\begin{equation}
\label{eq:eigenset-def}
z_i(S_k^{\pm}) = \pm T \delta_{ik}\;\;.
\end{equation}
We will extensively use the obtained basis sets $S_k^{\pm}$ and
Eq.~(\ref{eq:obserror-hessian}) in the remainder of the paper
to propagate uncertainties of FFs to several observables.

\subsection{Results}
%
\begin{figure}[!ht]
\begin{center}
\epsfig{figure=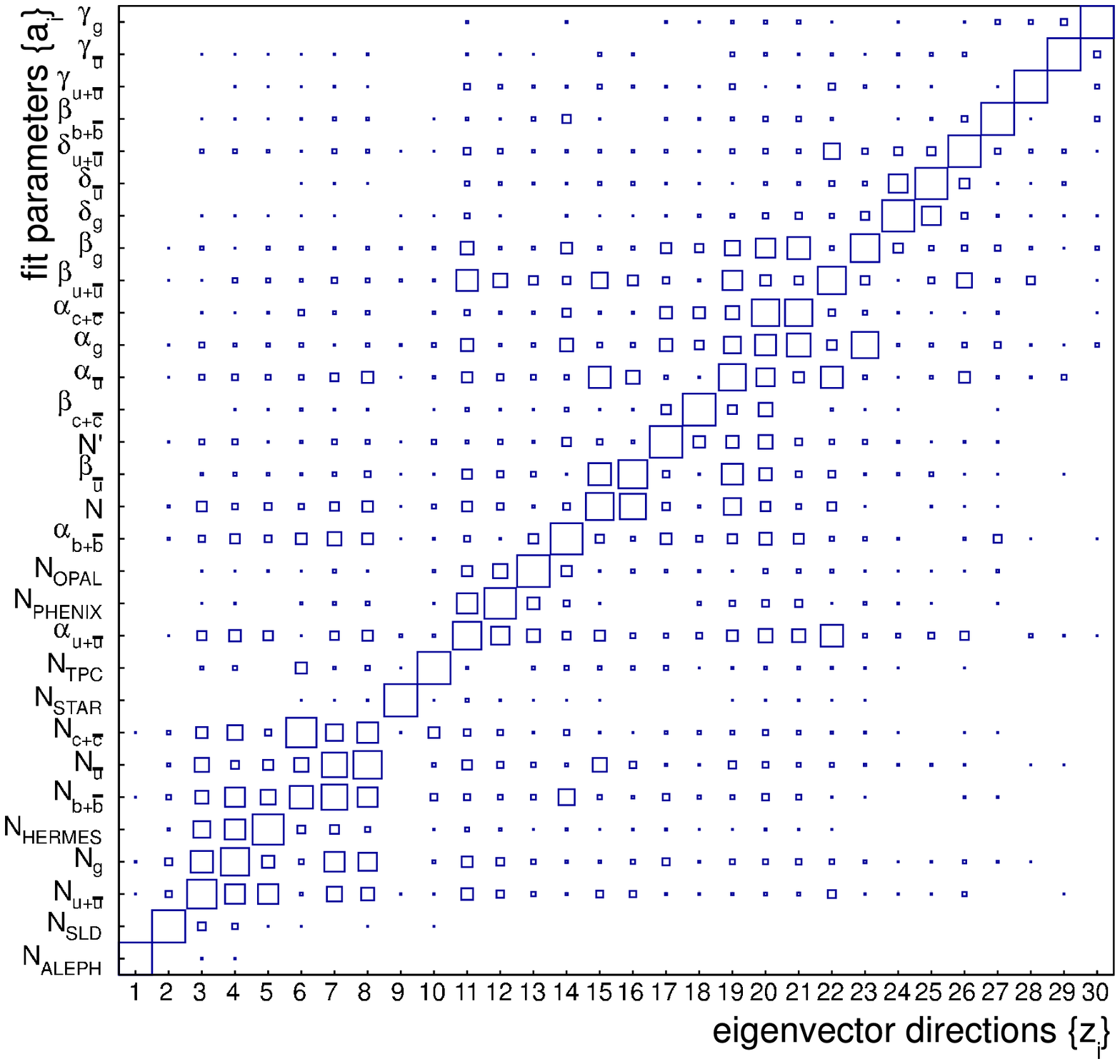,width=0.5\textwidth}
\end{center}
\vspace*{-0.5cm}
\caption{Correlations between the fit parameters $\{a_i\}$ 
and the eigenvector directions $\{z_i\}$. The larger the 
box size the larger the overlap, see text.
\label{fig:boxplot}}
\end{figure} 
First, before making use of the eigenvector sets, we need to validate 
the Hessian method as a reliable tool to determine uncertainties of pion FFs.
Figure~\ref{fig:boxplot} shows the overlap of each of the original fit parameters $\{a_i\}$ 
in Eq.~(\ref{eq:ff-input}) with the eigenvector directions $\{z_i\}$ introduced in
Sec.~\ref{sec:prel}. 
The larger the box size the larger the contribution
of a certain eigenvector direction to a fit parameter $a_i$.
The $\{z_i\}$ in Fig.~\ref{fig:boxplot} are ordered in terms of the size of the
corresponding eigenvalues  of the Hessian matrix: 
$z_1$ corresponds to the largest eigenvalue, i.e., the direction
in parameter space where $\chi^2$ changes most rapidly, whereas
$z_{30}$ is only very weakly constrained by data.
One can see that in most cases there is a fairly strong correlation between a 
given original fit parameter $a_i$ and a single eigenvector direction $z_j$. 
The opposite case, when several fit parameters are strongly correlated 
with an eigenvector direction, implies that those fit parameters are mutually 
correlated cannot be constrained independently. 

As can be inferred from Fig.~\ref{fig:boxplot}, the relative normalizations applied
to data sets in the fit, $\{N_{\mathrm{ALEPH}}$, $N_{\mathrm{SLD}}$, $N_{\mathrm{HERMES}}$,
$N_{\mathrm{STAR}}$, $N_{\mathrm{TPC}}$, $N_{\mathrm{PHENIX}}$, $N_{\mathrm{OPAL}}\}$, 
are among the best constrained parameters as they are typically linked to only a very small number 
of eigenvector directions, all corresponding to large eigenvalues.
The range of variation of these parameters is governed by the normalization uncertainty
quoted by each of the experiments and their main role is to ease possible tensions among the data
sets in the fit.

\begin{figure}[!th]
\begin{center}
\vspace*{-0.8cm}
\epsfig{figure=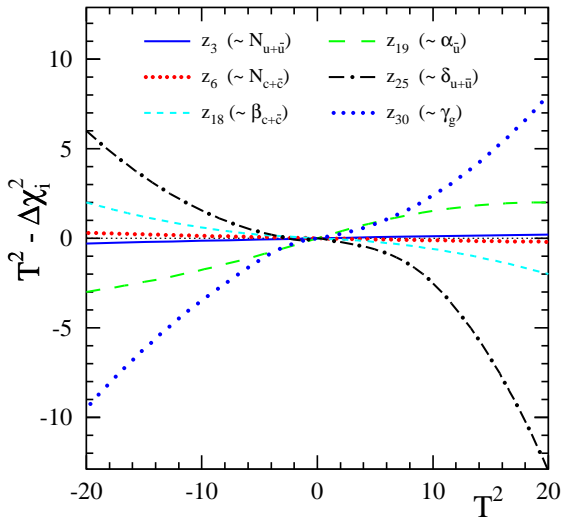,width=0.48\textwidth}
\end{center}
\vspace*{-0.8cm}
\caption{[color online] Examples of deviations from the expected parabolic behavior 
$\Delta \chi^2=T^2$ for selected eigenvector directions $\{z_i\}$, see text.
\label{fig:parabolicity}}
\end{figure}
In a second group of comparatively well constrained parameters 
one can find the normalization factors $N_i$,
see Eq.~(\ref{eq:ff-input}), of the different FFs for positively charged pions $D_i^{\pi^{+}}$, 
starting from the dominant (or favored) one, $N_{u+{\bar{u}}}$, 
followed by $N_{g}$ for the gluon FF, $N_{b+{\bar{b}}}$, $N_{\bar{u}}$, and $N_{c+{\bar{c}}}$. 
The $N_i$ represent the second moments of the FFs for flavor $i$ into hadron $H$,
\begin{equation}
\label{eq:secondmom}
D^H_i(Q^2) \equiv \int_{0}^1 z D_i^H(z,Q^2) dz, 
\end{equation}
entering the sum rule (\ref{eq:sumrule}).
One notices that the different $N_i$ are strongly correlated with each other.

The parameters $\alpha_i$ and $\beta_i$, controlling the main
features of the $z$ dependence in Eq.~(\ref{eq:ff-input}),
along with $N$ and $N^{\prime}$, related to certain
flavor symmetry relations among the FFs,
fall into the next category, all showing fairly large correlations.
Finally, the parameters associated with more subtle details of the $z$ dependence
of the FFs,
$\gamma_i$ and $\delta_i$, are the least well constrained ones in the fit
and are mainly correlated with eigenvector directions belonging to smaller eigenvalues.

Even though the correlations shown in Fig.~\ref{fig:boxplot} make explicit the hierarchy
of the fit parameters with respect to the level of how well they are constrained 
and to what extent correlations among them can be found, it does not necessarily
indicate if $\chi^2$ exhibits the {\em assumed} quadratic behavior on the
parameters away from the best fit.
To explore this further, Fig.~\ref{fig:parabolicity} illustrates 
the deviations of the $\chi^2$ function from the expected 
quadratic dependence for selected, representative 
eigenvector directions $\{z_i\}$.
Here, we vary only one of the parameters $z_i$ at a time such
that a given change of $\Delta\chi^2=T^2$ is produced.
Of course, since each $z_i$ has in principle overlap
with all fit parameters $\{a_i\}$, the latter all vary in this procedure.
For a truly quadratic behavior near the minimum, as is the underlying 
assumption in the Hessian approach, the quantity $T^2-\Delta\chi_i^2$ vanishes,
where $\Delta\chi_i^2$ is the change in $\chi^2$ induced by the variation
of the parameter $z_i$. Any deviation from zero will signal
a departure from the expected parabolic dependence.

As can be seen, for all the parameters $z_i$ shown in Fig.~\ref{fig:parabolicity}
the choice $T=1$, i.e., $\Delta\chi^2=1$, works very well, leading only to 
fairly small deviations from zero.
This implies that Hessian method is reliable for $\Delta\chi^2=1$, and 
our eigenvector sets $S_k^{\pm}$ are expected to produce faithful uncertainty estimates
close to those obtained with the robust LM approach. 
For some of the eigenvector directions $z_i$, those strongly constrained by data and corresponding to
large eigenvalues of the Hessian matrix, the quadratic behavior persists even far away from 
the $\chi^2$ minimum, i.e., up to large values of $T$.
Examples are $z_3$ and $z_6$, which are mainly correlated with 
the normalizations $N_{u+{\bar{u}}}$ and $N_{c+{\bar{c}}}$ of
the total up and charm quark FFs, respectively.
Some eigenvector directions, like, for instance, $z_{19}$, 
according to Fig.~\ref{fig:boxplot} mainly related to the small $z$
exponent $\alpha_{\bar{u}}$ of the anti-up-quark FF, and $z_{18}$,
controlling the large $z$ behavior of the total charm quark FF
through $\beta_{c+{\bar{c}}}$,
do show a more pronounced departure from the ideal parabolic behavior
starting already at about $T^2\simeq 5$.
Not surprisingly, deviations are most pronounced for poorly constrained
parameters such as $z_{25}$ and $z_{30}$ mainly correlated with
the dependence of total up quark and gluon FFs
at intermediate values of $z$ as described by 
$\delta_{u+{\bar{u}}}$ and $\gamma_g$, respectively.

\begin{figure}[!th]
\begin{center}
\vspace*{-0.6cm}
\epsfig{figure=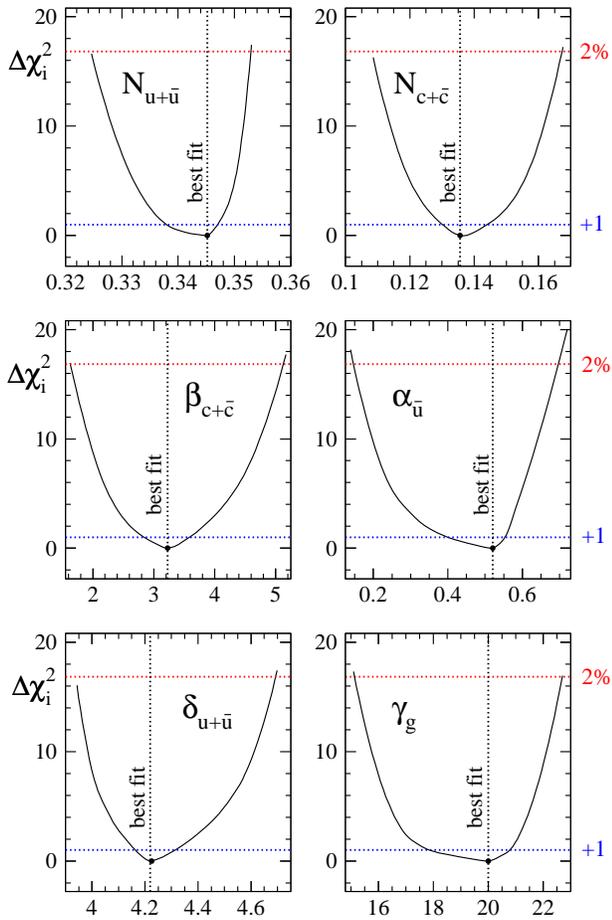,width=0.50\textwidth}
\end{center}
\vspace*{-0.9cm}
\caption{The $\chi^2$ profiles obtained with the LM approach
for the six fit parameters discussed in Fig.~\ref{fig:parabolicity}.
The horizontal lines indicate an increase $\Delta\chi^2_i$ 
by one unit or by $2\%$ of the total $\chi^2$ as adopted
in the DSS analysis \cite{deFlorian:2007aj}; see text.
\label{fig:par4}}
\vspace*{-0.5cm}
\end{figure}
The deviations from the assumed quadratic behavior observed above   
are also reflected in the actual $\chi^2$ profiles for the fit parameters
mainly associated with the eigenvector directions $z_i$ shown in 
Fig.~\ref{fig:parabolicity}. They are readily computed within the LM
technique and displayed in Fig.~\ref{fig:par4}.
Clearly, while for some of the parameters the profiles are reasonably smooth and parabolic,
as is assumed in the Hessian method, in general they are not. 
The profiles can exhibit asymmetric shapes, almost flat regions, and perhaps even
multiple minima. It is worth pointing out that in case of the DSS fit \cite{deFlorian:2007aj} none of these
features is related to a lack of flexibility in the chosen functional form,
Eq.~(\ref{eq:ff-input}). For instance, tensions among different data sets cause the 
asymmetry in the $\chi^2$ profile for $\alpha_{\overline{u}}$, and the flat part
in the profile for $\gamma_g$ is caused by insufficient constraints from data
in some kinematic regions.

At this point it is important to check whether the IH approach yields meaningful
uncertainty estimates for observables, despite the deviations from the assumed
quadratic dependence of $\chi^2$ on the parameters, illustrated in
Figs.~\ref{fig:parabolicity} and \ref{fig:par4}.
In the DSS analysis \cite{deFlorian:2007aj}, truncated second moments of the FFs
were identified as representative indicators of the typical uncertainties in the fit,
and their variations were studied within the LM framework.
They are defined in complete analogy to Eq.~(\ref{eq:sumrule}) by introducing a
lower cut-off $z_{\min}$ to the range of integration:
\begin{equation}
\label{eq:truncmom}
\eta^H_i(z_{\min},Q^2) \equiv \int_{z_{\min}}^1 z D_i^H(z,Q^2) dz\,. 
\end{equation}
To avoid the kinematic region of small $z$ where mass effects, neglected in
FFs, become relevant, $z_{\min}=0.2$ was chosen in \cite{deFlorian:2007aj}.
We note that the picture arising from the DSS analysis of the truncated second moments
$\eta^H_i(z_{\min},Q^2)$ \cite{deFlorian:2007aj} was recently shown to agree well 
with explicit calculations of uncertainties in hadron production cross sections 
at LHC kinematics based on the robust LM method \cite{Sassot:2010bh}.

\begin{figure*}[!thb]
\vspace*{-0.5cm}
\begin{center}
\epsfig{figure=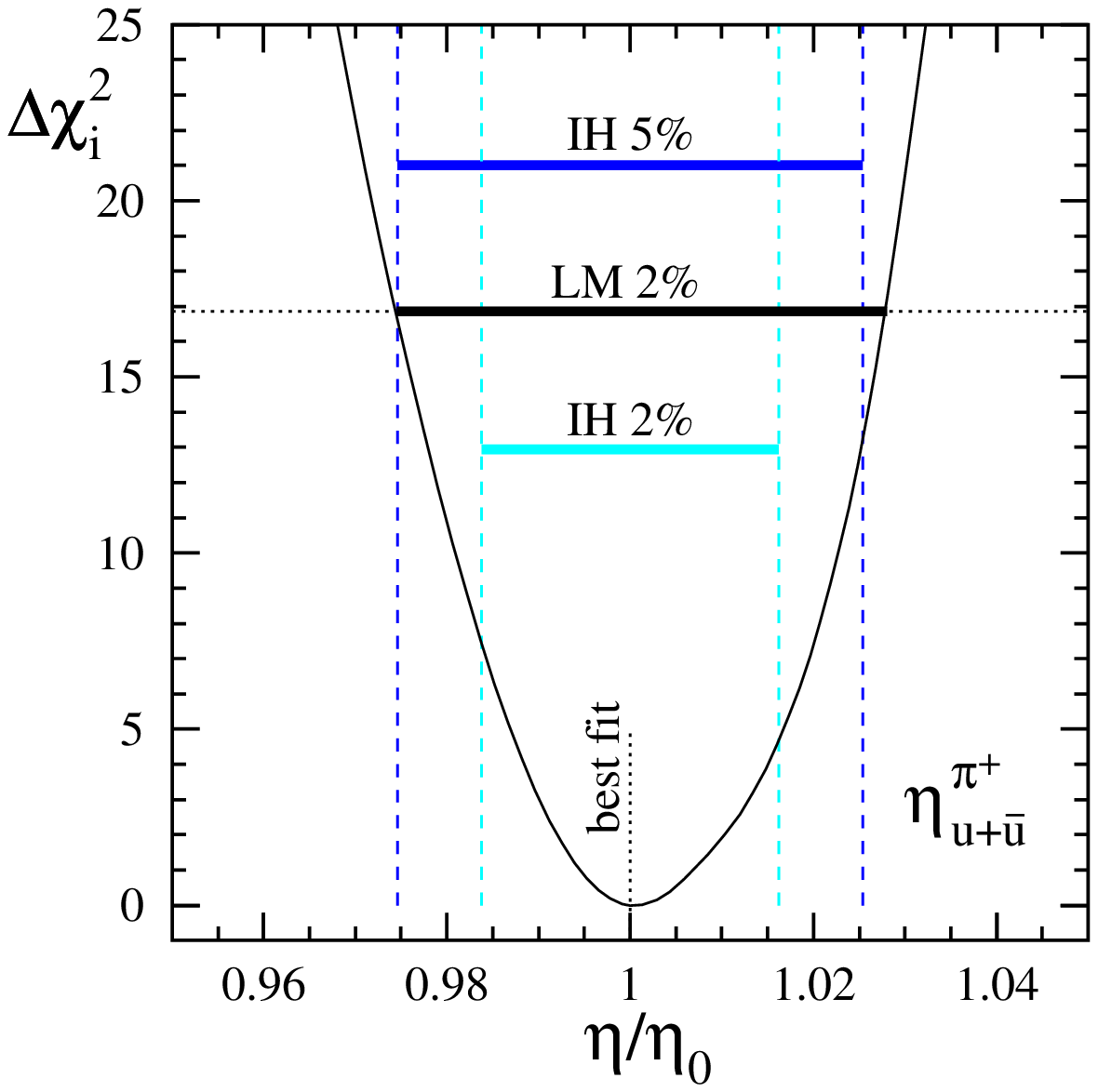,width=0.34\textwidth}
\hspace*{-0.6cm}
\epsfig{figure=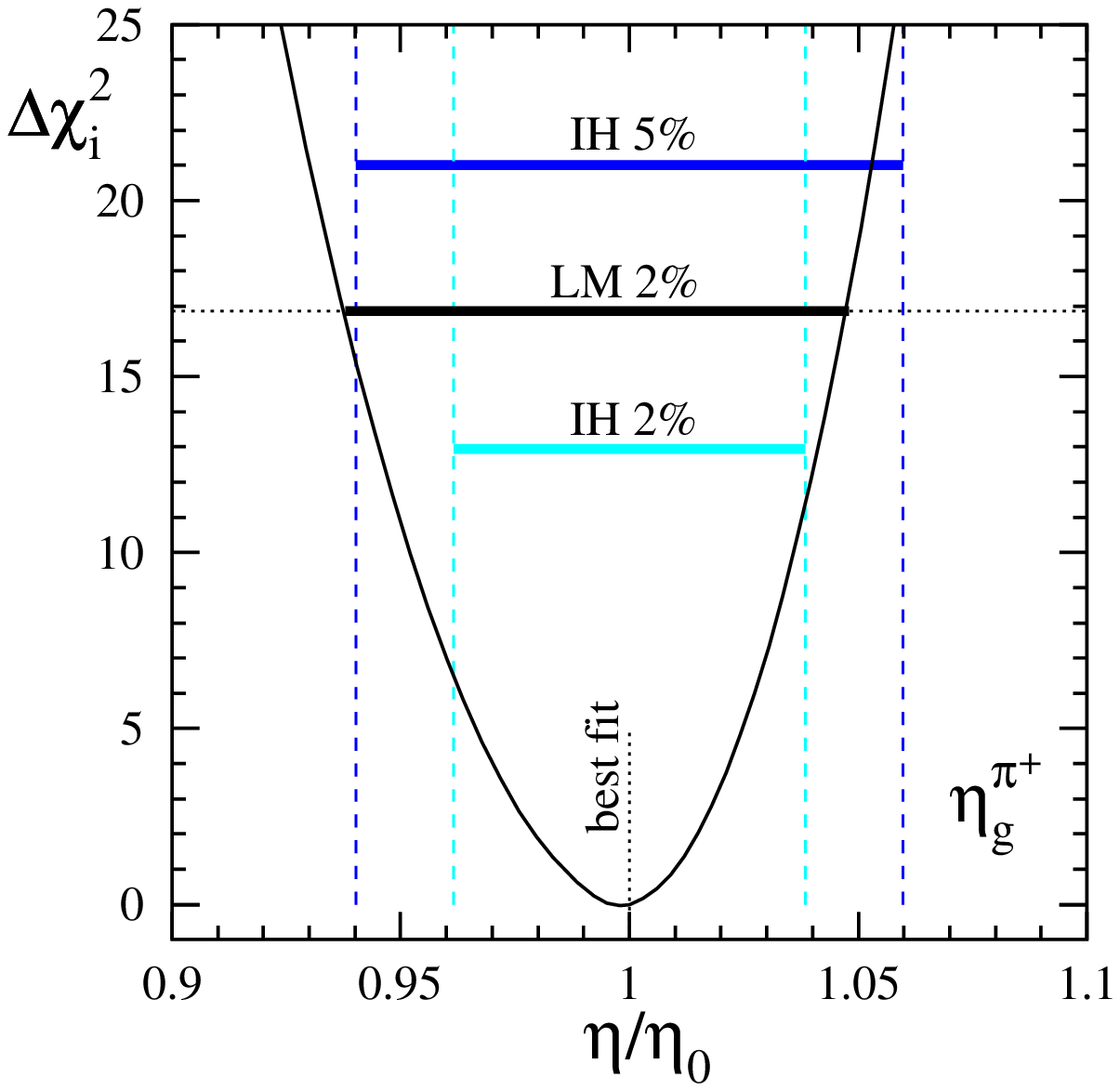,width=0.34\textwidth}
\hspace*{-0.6cm}
\epsfig{figure=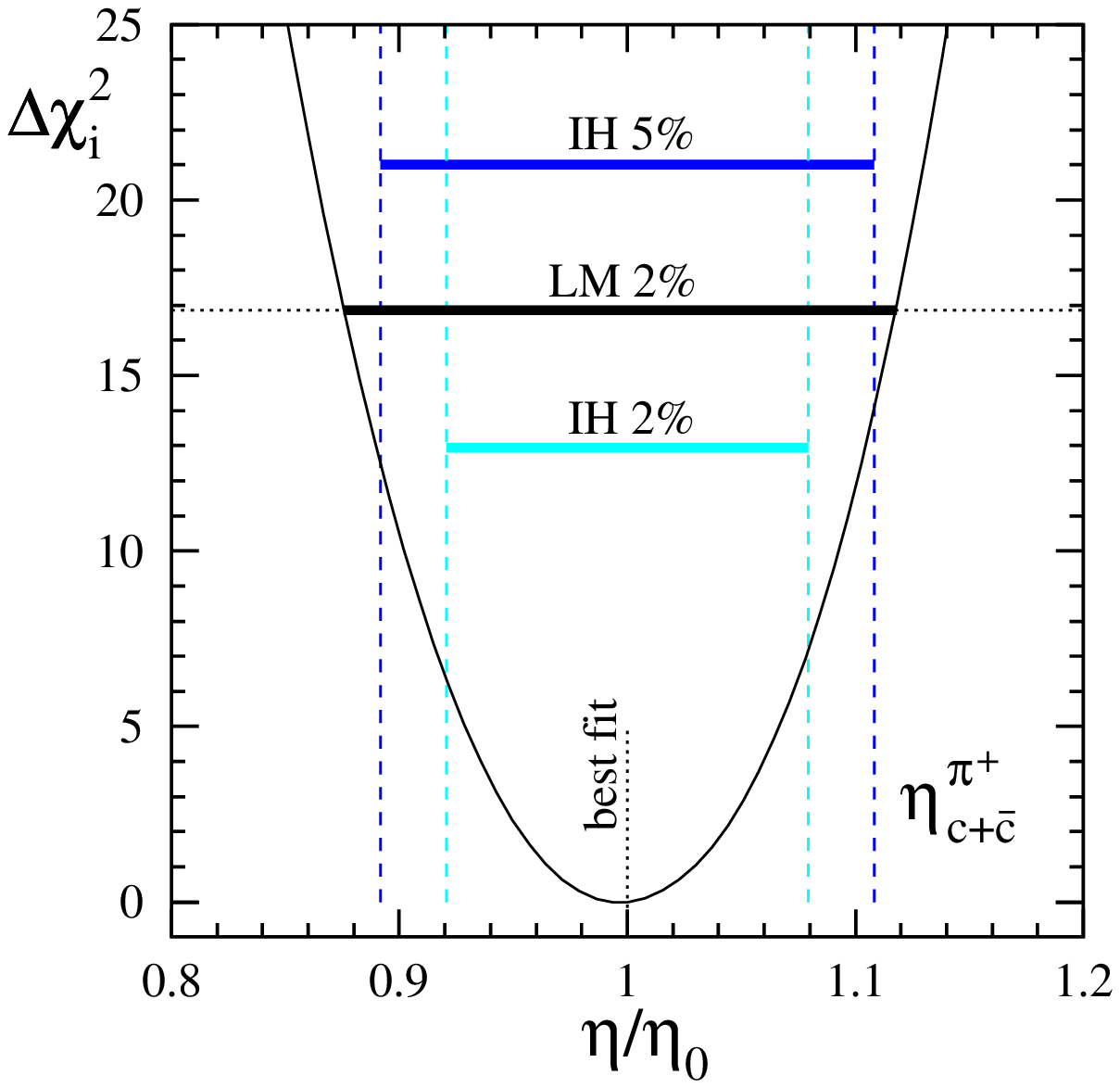,width=0.34\textwidth}
\end{center}
\vspace*{-0.6cm}
\caption{
Comparison of the uncertainties for selected truncated moments $\eta^{\pi^{+}}_i$
of the DSS pion FFs (indicated by the horizontal bars) 
estimated with the LM and IH methods at $Q=5\,\mathrm{GeV}$; see text. 
\label{fig:pionmom}}
\end{figure*}
In Figure \ref{fig:pionmom} we show the $\chi^2$-profiles (solid lines) for 
the truncated second moments $\eta^{\pi^+}_i$ of the DSS pion FFs
at NLO accuracy at $Q=5\,\mathrm{GeV}$ and for selected parton flavors 
$i=\{u+\bar{u},g,c+\bar{c}\}$ as obtained with the LM technique.
The moments are normalized to the value $\eta_0$ they take for the optimum fit
to data, characterized by the set of parameters $\{a_i^0\}$.
From the width of the curves at a given value of $\Delta \chi^2$,
one can read off the uncertainties on the various $\eta/\eta_0$, which can be in general asymmetric
with respect to the best fit, i.e., $\eta/\eta_0=1$.

In the DSS analysis of pion and kaon FFs \cite{deFlorian:2007aj}, a typical tolerance 
of $\Delta \chi^2/\chi^2=2\%$ was regarded as a faithful measure of uncertainties,
which amounts to an increase in the total $\chi^2$ of the fit to pion data by about
17 units as is indicated by the horizontal lines in Fig.~\ref{fig:pionmom}
labeled as ``LM $2\%$''. It is important to recall that this tolerance criterion 
is not derived from any theoretical argument but is an empirical estimate by requiring
that all data sets included in the global fit are still adequately described.
Such kind of choices at various levels of sophistication 
are typically made also in global QCD analyses of polarized and unpolarized parton densities,
see, e.g., \cite{deFlorian:2008mr,ref:pdf}. The naive criterion $\Delta \chi^2=1$ 
is usually regarded as too small to reliably account for PDF or FF uncertainties 
due to the complex nature of global fits, the different characteristics of the 
data sets, and various, often unaccounted sources of non-Gaussian theoretical errors.
As can be inferred from Fig.~\ref{fig:pionmom}, variations of the truncated second
moments are typically found to be of the order of $2\div 3\%$, $5\%$, and
around $10\%$ for the $u+\bar{u}$, gluon, and $c+\bar{c}$ to pion FFs.

Also in Fig.~\ref{fig:pionmom}, the horizontal lines labeled as ``IH $2\%$'' and ``IH $5\%$''
represent the same uncertainties as obtained with the LM method above but now are
estimated within the IH framework for two different tolerances 
(2 and 5 percent increase in the total $\chi^2$ of the fit, respectively).
Clearly, the $2\%$ criterion systematically underestimates the uncertainties
found in the robust LM approach. 
Not surprisingly, the differences have their origin in the assumed quadratic
behavior of the $\chi^2$ profile away from its minimum,
which, in general, is not fully adequate as we have already demonstrated above in Fig.~\ref{fig:parabolicity}.

Since the discrepancies between the LM and IH estimates based on $\Delta \chi^2/\chi^2=2\%$ 
are not large enough to completely invalidate the Hessian method, we devise a simple recipe 
to remedy its shortcomings in practical applications based on pion FFs.
We observe, that the IH method reproduces the LM results obtained 
for $\Delta \chi^2/\chi^2=2\%$ much better if one allows for a larger tolerance of $5\%$,
as is indicated by the horizontal lines for ``IH $5\%$'' in Fig.~\ref{fig:pionmom}. 
Hence, we propose to use preferably the IH method with $\Delta \chi^2/\chi^2=5\%$
to closely match the uncertainty estimates for pion FFs advocated in the DSS analysis 
\cite{deFlorian:2007aj} and provide relevant Hessian eigenvector basis sets $S_k^{\pm}$
for both a $\Delta \chi^2/\chi^2$ of $2\%$ and $5\%$ \cite{ref:sets}.

\begin{figure*}[htb!]
\begin{center}
\vspace*{-0.2cm}
\epsfig{figure=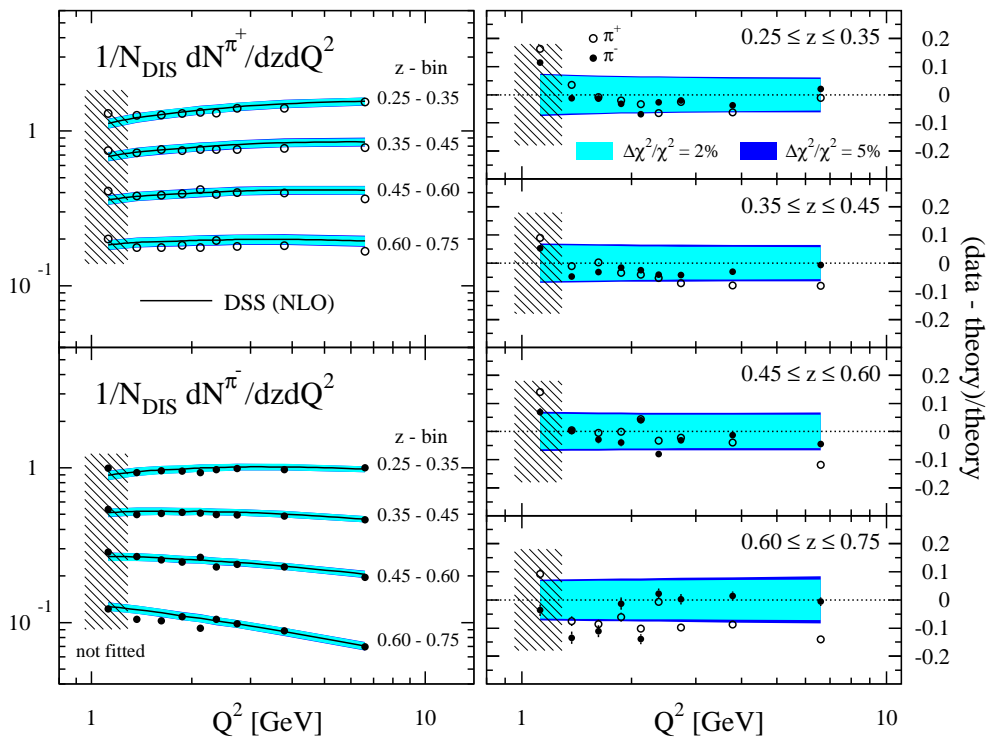,width=0.75\textwidth}
\end{center}
\vspace*{-0.7cm}
\caption{{\bf left:} comparison of the NLO DSS results for
charged pion multiplicities in SIDIS with preliminary data \cite{ref:hermessidis}
as a function of $Q^2$ in various bins of $z$.
{\bf right:} ``(data-theory)/theory'' for the DSS results; open and full
circles denote $\pi^+$ and $\pi^-$ multiplicities, respectively.
For all panels, the light and dark shaded bands indicate uncertainty estimates
for $\Delta \chi^2/\chi^2 = 2\%$ and $5\%$, respectively, based on
our Hessian eigenvector sets and using Eq.~(\ref{eq:obserror-hessian}).
\label{fig:sidis-pi}}
\vspace*{-0.5cm}
\end{figure*}
As an application and also to further verify the usefulness of the IH method,
we calculate uncertainty estimates for various representative data sets used in
the DSS analysis of pion FFs. 
To propagate the uncertainties of the FFs to pion production cross sections 
we use Eq.~(\ref{eq:obserror-hessian}) throughout.
Figure~\ref{fig:sidis-pi} illustrates the agreement between 
preliminary data for charged pion multiplicities \cite{ref:hermessidis}
in SIDIS and the corresponding results of the DSS fit at NLO accuracy \cite{deFlorian:2007aj}
in various bins of $z$. The results are largely independent of the choice of
PDFs. The multiplicity data are instrumental in providing flavor and charge
separated quark-to-pion fragmentation functions.
We notice that the differences between uncertainties obtained with
the $2\%$ and $5\%$ Hessian sets are less pronounced than in 
Fig.~\ref{fig:pionmom}, suggesting that the $2\%$ variations already account
for most of the error in the kinematic region probed by the multiplicity data.

\begin{figure*}[thb!]
\begin{center}
\vspace*{-0.6cm}
\epsfig{figure=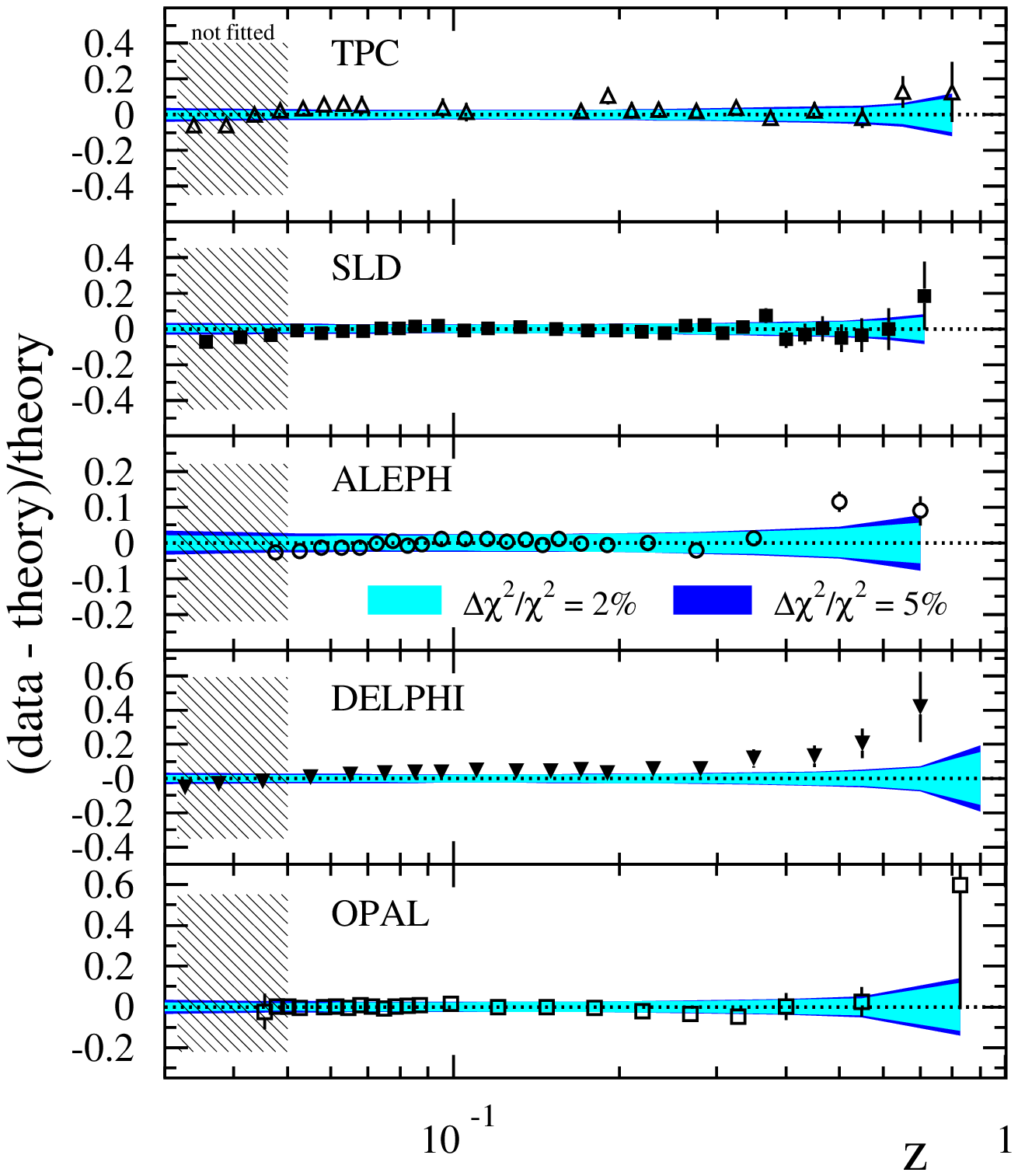,width=0.44\textwidth}
\epsfig{figure=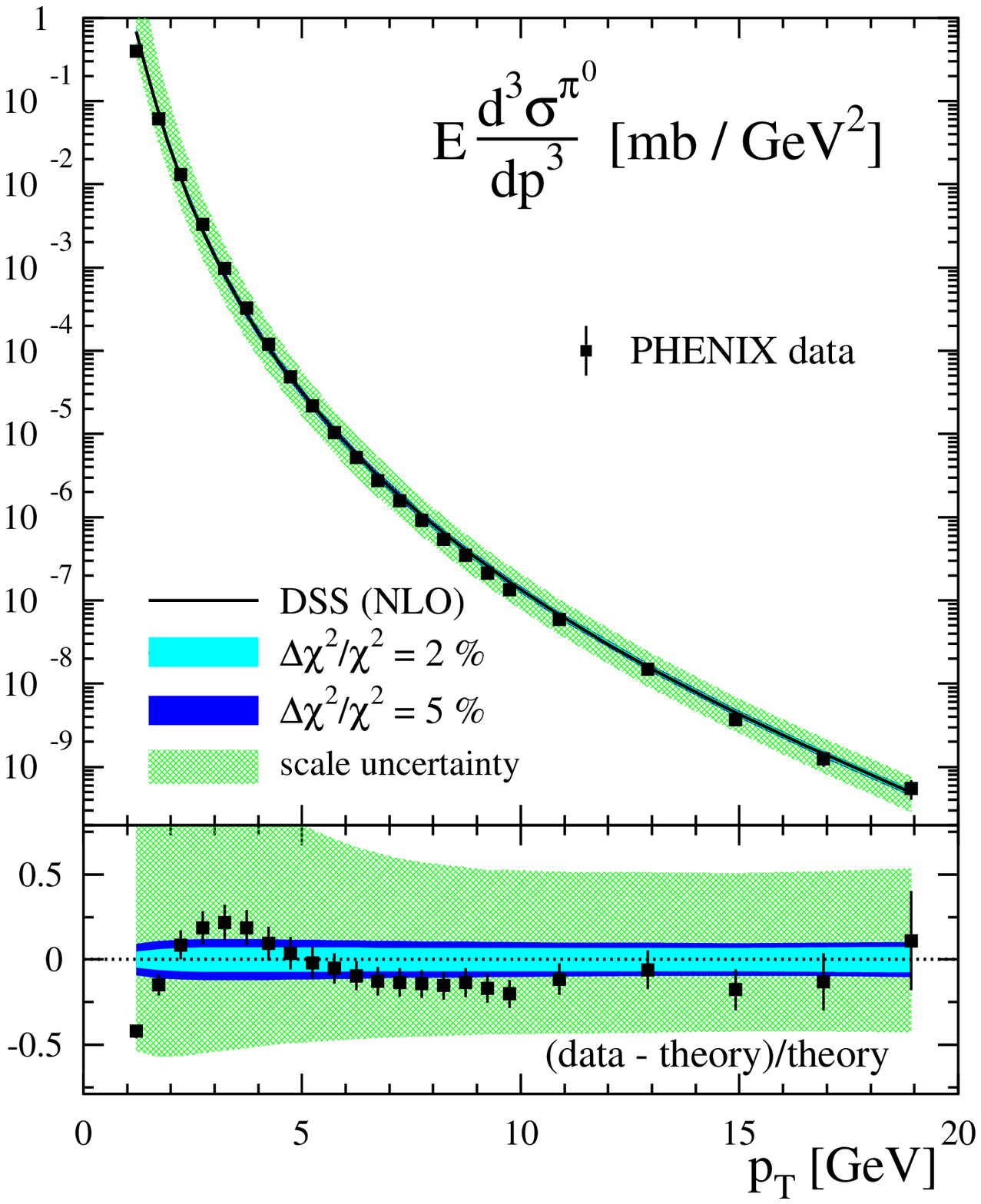,width=0.41\textwidth}
\end{center}
\vspace*{-0.7cm}
\caption{Uncertainty estimates for single-inclusive pion production in
SIA \cite{ref:sia-data} ({\bf left}) and in $pp$ collisions at $200\,\mathrm{GeV}$ 
and mid-rapidity \cite{ref:phenixpion} ({\bf right}). 
The light and dark shaded bands indicate the results
for $\Delta \chi^2/\chi^2 = 2\%$ and $5\%$, respectively, based on
our Hessian eigenvector sets and using Eq.~(\ref{eq:obserror-hessian})
For $pp\rightarrow \pi^0 X$, we also show the theoretical scale
ambiguity (outermost bands).
\label{fig:pion-sia-pp}}
\begin{center}
\vspace*{-0.7cm}
\epsfig{figure=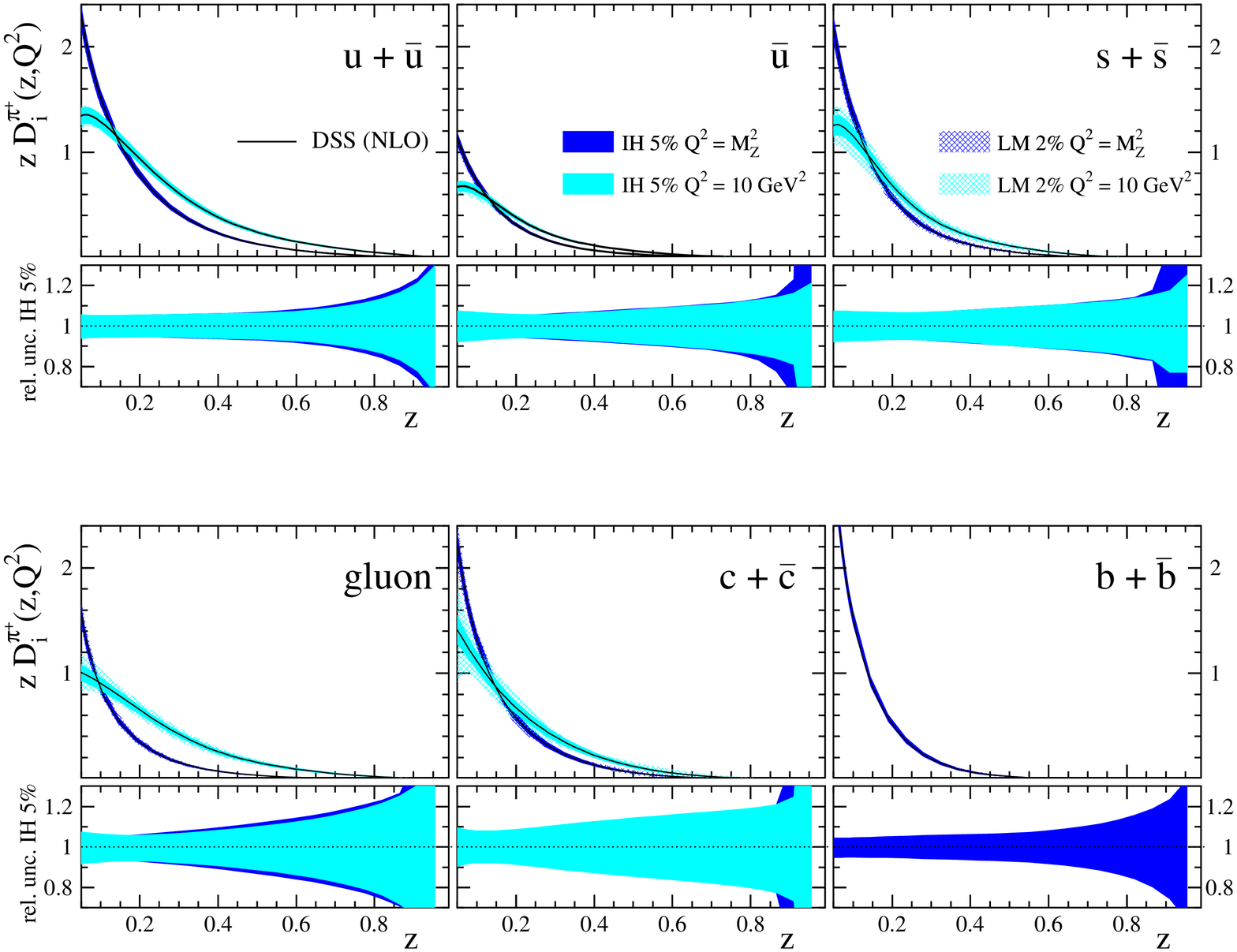,width=0.75\textwidth}
\end{center}
\vspace*{-2.cm}
\caption{Uncertainty bands for the DSS pion FFs estimated 
with the IH and LM methods at $Q^2=10\,\mathrm{GeV}^2$ and $Q^2=M_Z^2$. 
\label{fig:bandspion}}
\vspace*{-0.5cm}
\end{figure*}
Similar comparisons for single-inclusive pion production in SIA and $pp$ collisions at BNL-RHIC
are shown in Fig.~\ref{fig:pion-sia-pp}. Data are taken from 
\cite{ref:sia-data} and \cite{ref:phenixpion}, respectively.
Again, the differences between the uncertainty estimates obtained with the
2\% and 5\% Hessian eigenvector sets are minimal, except for regions
sensitive to large momentum fractions, $z\gtrsim 0.6$, where 
experimental constraints become very scarce. In these regions,
uncertainties are severely underestimated and not trustworthy.
In case of $pp$ collisions (right-hand-side of Fig.~\ref{fig:pion-sia-pp}),
we include for comparison an estimate of the theoretical ambiguity
due to the choice of the factorization scale in the NLO calculation 
\cite{ref:nlo} which is much more significant than errors on FFs (or PDFs 
as was shown in Ref.~\cite{Sassot:2010bh}). 

A common feature of the results shown in Figs.~\ref{fig:sidis-pi} 
and \ref{fig:pion-sia-pp} is that the estimates of the
relative uncertainties due to FFs remain almost constant in a wide range
of hadron momentum fractions $z$ and energy scale $\mu$ set
by the photon virtuality $Q$ in case of SIA or SIDIS 
or the transverse momentum $p_T$ in $pp$ collisions.
As expected, the obtained uncertainty bands cover and reflect 
the typical range of the statistical errors of the fitted data
relative to the best fit prediction.
We note that results very similar to those 
shown in Figs.~\ref{fig:sidis-pi} and \ref{fig:pion-sia-pp}
have been obtained for other observables depending on
pion FFs.

Finally, we address the uncertainty estimates on the individual
parton-to-pion FFs. Figure~\ref{fig:bandspion} shows the NLO DSS
$zD_i^{\pi^{+}}(z,Q^2)$ for $i=u+\bar{u}$, $\bar{u}$, $s+\bar{s}$,
$g$, $c+\bar{c}$, and $b+\bar{b}$ for two different scales
$Q^2=10\,\mathrm{GeV}^2$ and $Q^2=M_Z^2$ along with our estimates
of their uncertainties using both the IH and LM method.
For better visibility and to facilitate comparisons between the different FFs,
the lower panels for each flavor show the relative uncertainties
which are typically of the order of $5\%$ for the
favored quark combination $u+\bar{u}$ and around $10\%$ for
unfavored quark-to-pion and gluon FFs, fairly independent of
the scale $Q$. Uncertainties
increase significantly for large momentum fractions, $z\gtrsim 0.6$, where
current experimental constraints are insufficient.

In the DSS analysis \cite{deFlorian:2007aj}, the fragmentation of charm and bottom quarks into
charged pions is included discontinuously as massless partons
in the scale evolution above their $\overline{\text{MS}}$ 
``thresholds'', $Q=m_{c,b}$, with $m_{c,b}$ denoting the mass 
of the charm and bottom quark, respectively.
This simplified treatment of heavy flavors is at variance
with current PDF fits \cite{ref:pdf,Alekhin:2011sk}, where very elaborate schemes have been
developed to properly include mass effects near threshold and to
resum potentially large logarithms for $Q^2\gg m_{c,b}^2$.
Only flavor-tagged SIA data constrain the charm and bottom FFs
in the DSS analysis, and their uncertainties, albeit reasonably
small, should be taken with a grain of salt.

Concerning the comparison between the two methods to obtain
our uncertainty estimates, we find good agreement 
if the IH eigenvector sets with the $5\%$ tolerance criterion 
for $\Delta \chi^2/\chi^2$ are used. Only for $z\lesssim 0.3$ and unfavored,
i.e., less well constrained FFs, the LM method yields somewhat
larger uncertainties compared to the corresponding bands obtained with the IH technique
as can be seen, for instance in the panels showing the  
$\bar{u}$, $s+\bar{s}$, and the gluon FFs.

While the robust LM method is straightforwardly applied to cross sections
or truncated moments of FFs, it is not very handy in
determining the $z$ dependent uncertainties.
One can make use, for instance, of the profiles shown in Fig.~\ref{fig:par4} 
for the variations of each fit parameter $\{a_i\}$ 
to compute the spread in the FFs for each parton flavor $i$.
In doing so, one neglects, however, correlations among the
variations of the fit parameters, which can be non-negligible as we
have demonstrated in Fig.~\ref{fig:boxplot}. 
This might lead to an overestimate of uncertainties for the
$z$ dependent FFs. 
On the other hand, the Hessian method is 
particularly simple, using Eq.~(\ref{eq:obserror-hessian}), provided 
its applicability has been carefully established and appropriate eigenvector
sets have been generated \cite{ref:sets}. 
This was the purpose of our studies for pion FFs presented in this Section.

\section{Kaon Fragmentation Functions}
\subsection{Preliminaries}
Next, we proceed with a similar error analysis for the NLO DSS kaon
FFs \cite{deFlorian:2007aj}. Since even the most precise kaon production data
in SIA exhibit experimental uncertainties typically at least twice as large
as those found for pions, one must expect much less, but still 
reasonably well constrained parton-to-kaon FFs.

To account for the phenomenological expectation that the formation of secondary $s\bar{s}$ pairs
should be suppressed in the production of, say, a $|K^+\rangle=|u\bar{s}\rangle$,
the two favored quark combinations $D_{u+\bar{u}}^{K^+}$ and
$D_{s+\bar{s}}^{K^+}$ are fitted independently in the DSS analysis using the
functional form (\ref{eq:ff-input}). Indeed, in line with that expectation,
the DSS fit prefers $D_{s+\bar{s}}^{K^+}> D_{u+\bar{u}}^{K^+}$. 
Since presently available data do not fully constrain all unfavored kaon FFs,
is was assumed that they all share the same functional form, i.e.,
\begin{equation}
\label{eq:sea_ka}
D_{\bar{u}}^{K^+}=D_{s}^{K^+}=D_d^{K^+}=D_{\bar{d}}^{K^+}.
\end{equation}
Corresponding FFs into $K^-$ are obtained, as usual, 
by charge conjugation symmetry, which leaves a total of 24 free fit parameters 
$\{a_i\}$ describing the DSS FFs for quarks and gluons into positively
charged kaons. Six additional parameters control the relative normalization
of the data sets in the DSS analysis \cite{deFlorian:2007aj}.

\subsection{Results}
\begin{figure}[!bht]
\begin{center}
\epsfig{figure=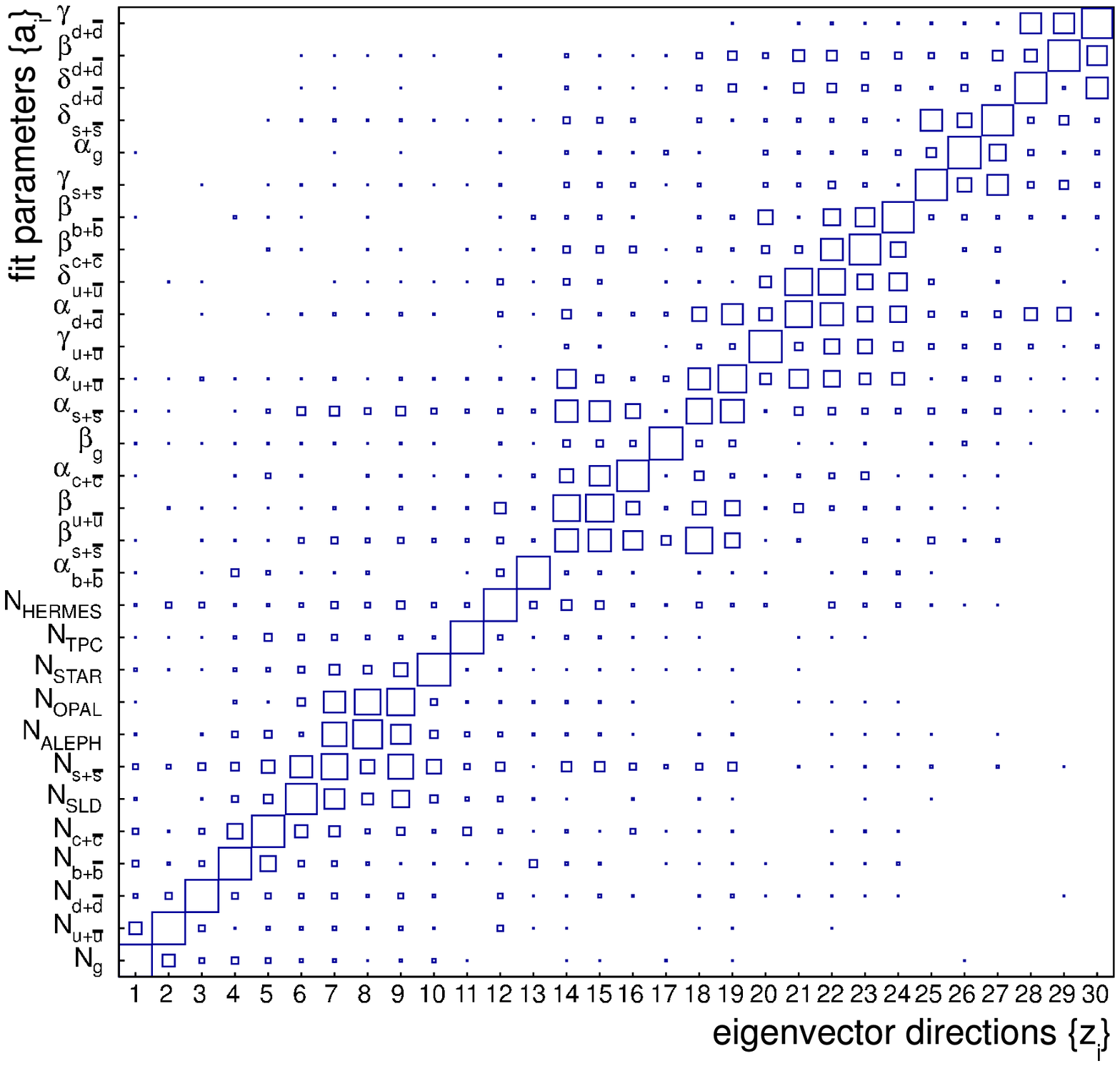,width=0.5\textwidth}
\end{center}
\vspace*{-0.5cm}
\caption{As in Fig.~\ref{fig:boxplot} but now showing
the correlations between the fit parameters $\{a_i\}$ 
and the eigenvector directions $\{z_i\}$ for positively charge 
kaons. 
\label{fig:boxplotka}}
\end{figure}

%
As for the pion FFs, we start our analyses by showing the correlations between the fit
parameters $\{a_i\}$ and eigenvector directions $\{z_i\}$
in Fig.~\ref{fig:boxplotka}.
As before, the $\{z_i\}$ are ordered by the size of the 
corresponding eigenvalues of the Hessian matrix, and correlations
can be found across the entire parameter space. 
In most cases, there are, however, fairly strong correlations
relating a certain eigenvector direction with just a one (or only a few) 
of the original fit parameters.
Again, the best constrained fit parameters are those related to relative
normalizations applied to the data sets in the fit. In a second group are
the normalization factors $N_i$ of the different FFs for positively charged
kaons $D_i^{K^{+}}$, see Eq.~(\ref{eq:ff-input}).
Among the least well constrained parameters are mainly those associated with
subtle details of the $z$ dependence of the FFs, i.e., $\gamma_i$ and $\delta_i$ 
in Eq.~(\ref{eq:ff-input}).
%
%

\begin{figure}[!th]
\begin{center}
\vspace*{-0.75cm}
\epsfig{figure=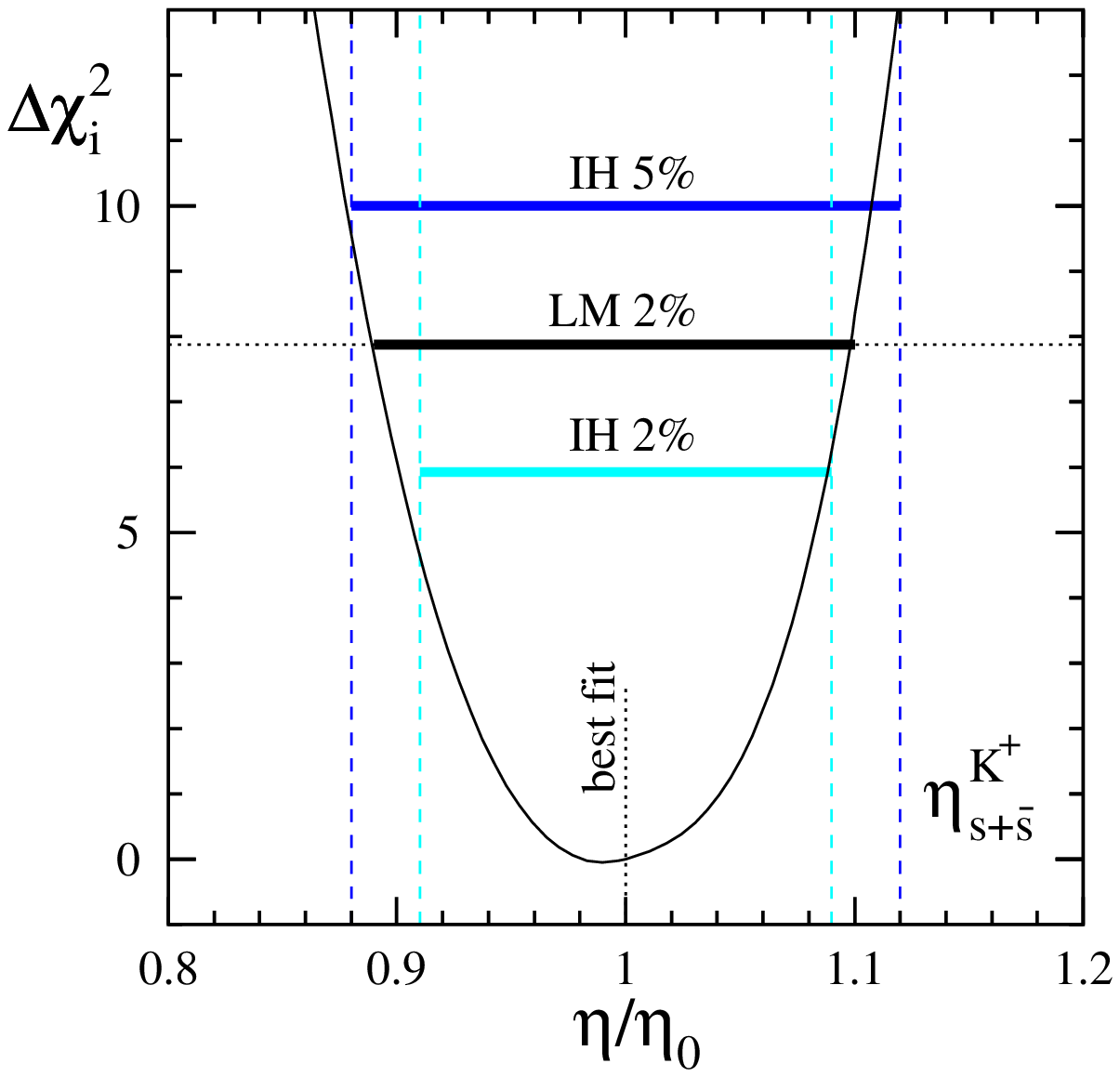,width=0.48\textwidth}
\end{center}
\vspace*{-0.6cm}
\caption{As in Fig.~\ref{fig:pionmom} but now comparing the range of variation
(indicated by the horizontal bars) of the truncated moment $\eta^{K^{+}}_i$ 
of the dominant  $(s+\bar{s})$-to-kaon FFs at $Q=5\,\mathrm{GeV}$; see text.
\label{fig:kaonmom}}
\end{figure}
In the DSS analysis of kaon FFs \cite{deFlorian:2007aj}, uncertainties were 
estimates within the LM method for the truncated moments (\ref{eq:truncmom}) 
and found to be at least twice as large as those for pions. 
Again, an increase of $\Delta \chi^2/\chi^2=2\%$, corresponding to about 8 units
in $\chi^2$, were regarded as a faithful estimate of the typical uncertainties.
As a representative example,
Fig.~\ref{fig:kaonmom} shows the corresponding $\chi^2$ profile (solid line)
and error estimate (horizontal bar labeled as ``LM $2\%$'') for the dominant total strange quark
FFs into positively charged kaons. As in Fig.~\ref{fig:pionmom},
the variation of $\eta_{s+\bar{s}}^{K^{+}}$ is normalized to its value $\eta_0$
taken for the optimum set of parameters $\{a_i^0\}$ and is found to be
of order of about $10\%$.

When compared to the uncertainties obtained within the IH approach,
we find that the results corresponding to $\Delta \chi^2/\chi^2=2\%$
and $5\%$ slightly under- and overestimate the range of variation of
$\eta_{s+\bar{s}}^{K^{+}}$ determined with the LM method.
Qualitatively very similar results can be obtained for all 
the other parton-to-kaon FFs.
Again, we provide Hessian eigenvector basis sets $S_k^{\pm}$ for
both tolerance criteria \cite{ref:sets} to facilitate 
estimates of uncertainties for arbitrary observables involving
kaon FFs.

\begin{figure*}[ht]
\begin{center}
\vspace*{-0.7cm}
\epsfig{figure=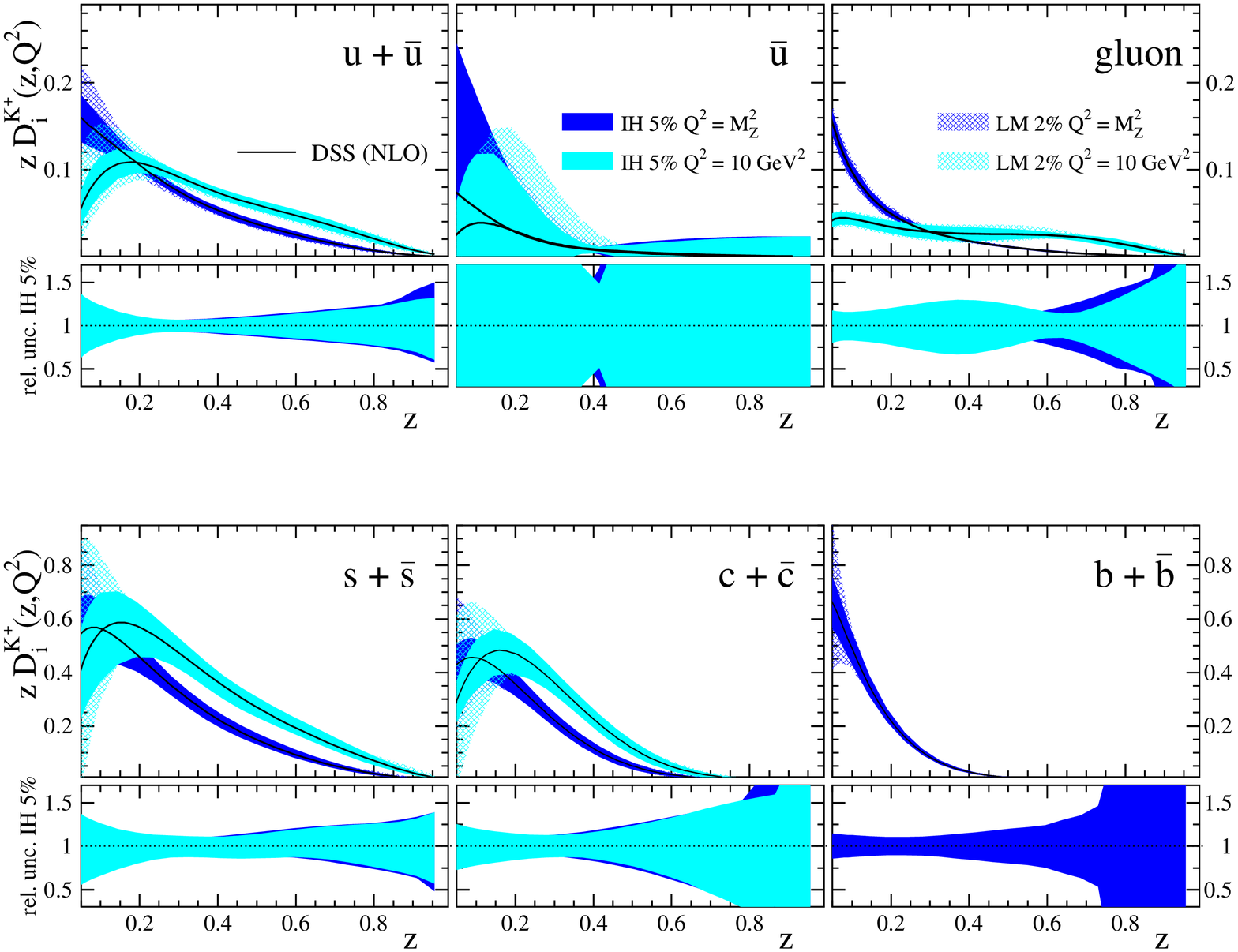,width=0.8\textwidth}
\end{center}
\vspace*{-2.cm}
\caption{As in Fig.~\ref{fig:bandspion} but now showing the uncertainty 
bands for the NLO DSS kaon FFs at $Q^2=10\,\mathrm{GeV}^2$ and $Q^2=M_Z^2$. 
\label{fig:bandskaon}}
\vspace*{-0.5cm}
%
\begin{center}
\vspace*{-0.3cm}
\epsfig{figure=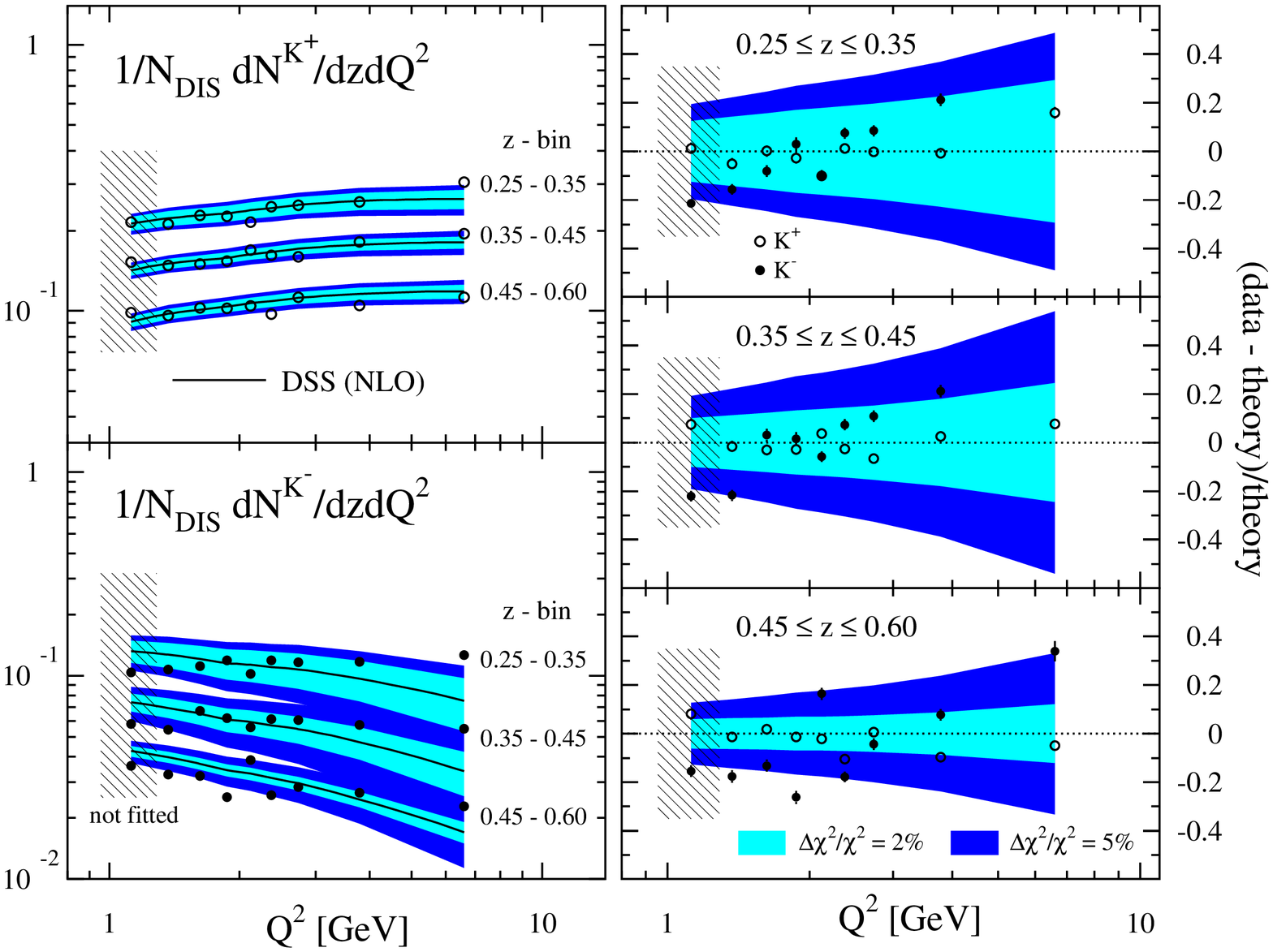,width=0.75\textwidth}
\end{center}
\vspace*{-0.7cm}
\caption{As in Fig.~\ref{fig:sidis-pi} but for charged kaon multiplicities.
On the right-hand-side, only the uncertainty bands for $K^-$ are visible, which
are larger than those for $K^+$.
\label{fig:sidis-ka}}
\vspace*{-0.5cm}
\end{figure*}

Before discussing an application of our Hessian sets, we first study
the uncertainty bands for the individual NLO DSS parton-to-kaon
FFs. Figure~\ref{fig:bandskaon} shows $zD_i^{K^{+}}(z,Q^2)$ 
for $i=u+\bar{u}$, $\bar{u}$, $g$, $s+\bar{s}$,
$c+\bar{c}$, and $b+\bar{b}$ for two different scales
$Q^2=10\,\mathrm{GeV}^2$ and $Q^2=M_Z^2$ along with our estimates
of their uncertainties using both the IH ($5\%$) and LM ($2\%$) method.
As for pion FFs, the lower panels for each flavor show the 
relative uncertainties obtained with the IH approach.
 
Despite the sizable disparity in size, the two favored
$s+\bar{s}$ and $u+\bar{u}$ FFs exhibit very similar relative uncertainties: 
close to $20$\% at intermediate values of $z$ and 
a significant growth for both large and small $z$.
All unfavored quark FFs, such as $D_{\bar{u}}^{K^{+}}$,
are not well constrained by data and show extremely large uncertainties 
of $50\%$ or more.
All these features can be traced back to the much larger experimental
uncertainties for flavor-tagged SIA \cite{ref:sia-data} and SIDIS \cite{ref:hermessidis}
data for kaons compared to those for identified pions.

Rather unexpectedly, the gluon-to-kaon FF shows rather small relative uncertainties 
over most of the $z$-range, comparable to those for the favored quark FFs.
This feature is found both within the LM and IH method and hence cannot
be attributed to the approximations made in the latter approach.
Since the $pp$ data, which mainly constrain $D_g^{K^{+}}$ at $z\simeq 0.5$, 
are rather sparse and less precise than for identified pions, 
the result most likely reflects a lack of flexibility in the parametrization
chosen in the DSS analysis, albeit not affecting the overall quality of the fit.
More and better data will be needed to constrain $D_g^{K^{+}}$ more reliably.
In particular, upcoming precise data from BELLE \cite{ref:belleprel} 
might be sufficient to determine the gluon FF from scaling violations in SIA.
Additional $pp$ data from RHIC and the LHC will add to this.
We note that as in the case of pions, charm and bottom quark FFs and their uncertainties
have to taken with a grain of salt.
  
In order to illustrate the impact of the estimated uncertainties for the DSS kaon FFs, we compute
the charged kaon multiplicities at NLO accuracy in the kinematic range of the
HERMES experiment \cite{ref:hermessidis}. To propagate the uncertainties of the kaon FFs we use 
Eq.~(\ref{eq:obserror-hessian}) with our Hessian sets for both $\Delta \chi^2/\chi^2=2\%$ and $5\%$.
The results are shown in Fig.~\ref{fig:sidis-ka}.
For positively charged kaons, the uncertainties resemble in shape those found 
in the case of pions, see Fig.~\ref{fig:sidis-pi}, i.e., they are largely independent
of $z$ and the energy scale $Q$. As has to be expected from a comparison of 
Figs.~\ref{fig:bandspion} and \ref{fig:bandskaon}, the differences between results obtained
with the Hessian sets for the $2\%$ and $5\%$ tolerances are more
pronounced for kaons than for pions.
For negatively charged kaon multiplicities the propagated uncertainties behave, however,
rather differently. We observe an increase in size with the scale $Q$, and,
in addition, the estimates for the $2\%$ and $5\%$ Hessian sets differ much more significantly
than for $K^+$ and $\pi^{\pm}$ multiplicities.
These features can be associated with the larger sensitivity
of the $K^-$ multiplicities to the less well determined 
unfavored kaon FFs as well as with the more erratic behavior of the fitted SIDIS data. 

We note that kaon FFs are extremely relevant in connection with the proper extraction
of the strangeness helicity PDF from spin-dependent SIDIS data 
\cite{deFlorian:2008mr,Leader:2010rb}.
Our error estimates for pion and kaon FFs will be particularly useful in future global 
analyses of the spin structure of the nucleon as they allow one to straightforwardly 
propagate these sources of uncertainties.
Upcoming new data on multiplicities for identified hadrons \cite{ref:compassmult} 
will be critical in further reducing uncertainties of fully flavor and charge separated sets of FFs.

\section{Conclusions}
%
We have performed a detailed estimate of the uncertainties inherent to the
extraction of pion and kaon fragmentation functions as performed in 
the DSS global analysis framework.
We have carefully examined the validity of the approximations underlying the
more practical, improved Hessian method by comparing with the results
based on the robust Lagrange multiplier technique.

Even though we have found some differences between uncertainties obtained with 
the two methods, they can be readily understood and accounted for by choosing
are larger tolerance in the Hessian approach. In general, the agreement is
much better for the rather well constrained parton-to-pion fragmentation functions
and somewhat less satisfactory for kaons, where data are less precise.

We provide Hessian eigenvector basis sets for both pion and kaon fragmentation functions
and corresponding to two different error estimates \cite{ref:sets}.
These sets will greatly facilitate the propagation of uncertainties related to
fragmentation functions to observables such as single-inclusive hadron production
cross sections or multiplicities. Good knowledge of fragmentation functions
and their uncertainties is also relevant in understanding and analyzing results
for spin-dependent semi-inclusive deep-inelastic scattering and the
modification of identified hadron yields in heavy-ion collisions.
Our results will also prove to be relevant in quantifying the impact of future
measurements of pion and kaon yield in upcoming global analyses of 
fragmentation functions. 

\section*{Acknowledgments}
%
M.S.\ acknowledges support by the U.S.\ Department of Energy under contract 
number DE-AC02-98CH10886 and a BNL ``Laboratory Directed Research and Development'' grant 
(LDRD 12-034).
This work was partially supported by CONICET, ANPCyT, and UBACyT.



\begin{thebibliography}{99}
%
\bibitem{deFlorian:2008mr} 
  D.~de Florian, R.~Sassot, M.~Stratmann, and W.~Vogelsang,
  Phys.\ Rev.\ Lett.\  {\bf 101}, 072001 (2008);
  Phys.\ Rev.\ D {\bf 80}, 034030 (2009).
%
\bibitem{Leader:2010rb} 
  E.~Leader, A.~V.~Sidorov, and D.~B.~Stamenov,
  Phys.\ Rev.\ D {\bf 82}, 114018 (2010).
%
\bibitem{Sassot:2009sh} 
  R.~Sassot, M.~Stratmann, and P.~Zurita,
  Phys.\ Rev.\ D {\bf 81}, 054001 (2010).
%
\bibitem{deFlorian:2011fp} 
  D.~de Florian, R.~Sassot, P.~Zurita, and M.~Stratmann,
  Phys.\ Rev.\ D {\bf 85}, 074028 (2012).
%
\bibitem{ref:kretzer} S.\ Kretzer, Phys. Rev. {\bf D62}, 054001 (2000).
%
\bibitem{ref:kkp} B.A.\ Kniehl, G.\ Kramer, and B.\ P\"{o}tter,
Nucl. Phys. {\bf B582}, 514 (2000).
%
\bibitem{ref:akk} S.\ Albino, B.A.\ Kniehl, and G.\ Kramer,
Nucl. Phys. {\bf B725}, 181 (2005).
%
\bibitem{Hirai:2007cx} 
  M.~Hirai, S.~Kumano, T.~-H.~Nagai, and K.~Sudoh,
  Phys.\ Rev.\ D {\bf 75}, 094009 (2007).
%
\bibitem{deFlorian:2007aj} 
  D.~de Florian, R.~Sassot, and M.~Stratmann,
  Phys.\ Rev.\ D {\bf 75}, 114010 (2007).
%
\bibitem{de Florian:2007hc} 
  D.~de Florian, R.~Sassot, and M.~Stratmann,
  Phys.\ Rev.\ D {\bf 76}, 074033 (2007).
%
\bibitem{Albino:2008fy} 
  S.~Albino, B.~A.~Kniehl, and G.~Kramer,
  Nucl.\ Phys.\ B {\bf 803}, 42 (2008).
%
\bibitem{ref:pdf} 
  A.~D.~Martin, W.~J.~Stirling, R.~S.~Thorne, and G.~Watt,
  Eur.\ Phys.\ J.\ C {\bf 63}, 189 (2009);
  H.~-L.~Lai, M.~Guzzi, J.~Huston, Z.~Li, P.~M.~Nadolsky, J.~Pumplin, and C.~-P.~Yuan,
  Phys.\ Rev.\ D {\bf 82}, 074024 (2010).
%
\bibitem{Albino:2008aa} 
  For an overview of recent analyses of FFs, see, e.g., S.~Albino {\it et al.},
  {\tt arXiv:0804.2021}.
%
\bibitem{Alekhin:2011sk} 
  For an overview of recent PDF sets and uncertainty estimates, see, e.g.,
  S.~Alekhin {\it et al.},
  ``The PDF4LHC Working Group Interim Report'', {\tt arXiv:1101.0536}.
%
\bibitem{Stump:2001gu} 
  D.~Stump {\it et al.},
  Phys.\ Rev.\ D {\bf 65}, 014012 (2001).
%
\bibitem{Pumplin:2001ct} 
  J.~Pumplin {\it et el.},
  Phys.\ Rev.\ D {\bf 65}, 014013 (2001);
  J.~Pumplin, D.~R.~Stump, and W.~K.~Tung,
  Phys.\ Rev.\  D {\bf 65}, 014011 (2002).    
%
\bibitem{Ball:2010de} 
  R.~D.~Ball, L.~Del Debbio, S.~Forte, A.~Guffanti, J.~I.~Latorre, J.~Rojo, and M.~Ubiali,
  Nucl.\ Phys.\ B {\bf 838}, 136 (2010).
%
\bibitem{ref:belleprel} M.\ Leitgab, talk presented at the ``XX Int.\ Workshop on
Deep-Inelastic Scattering and Related Subjects'', March 2012, Bonn, Germany.
%
\bibitem{ref:compassmult} N.\ Makke, talk presented at the ``XX Int.\ Workshop on
Deep-Inelastic Scattering and Related Subjects'', March 2012, Bonn, Germany.
%
\bibitem{Agakishiev:2011dc} 
  G.~Agakishiev {\it et al.} [STAR Collaboration],
  Phys.\ Rev.\ Lett.\  {\bf 108}, 072302 (2012).
%
\bibitem{Abelev:2012cn} 
  B.~Abelev {\it et al.} [ALICE Collaboration],
  {\tt arXiv:1205.5724}.
%
\bibitem{Sassot:2010bh} 
  R.~Sassot, P.~Zurita, and M.~Stratmann,
ÊÊPhys.\ Rev.\ D {\bf 82}, 074011 (2010).
ÊÊ
%
\bibitem{ref:sets} Hessian eigenvector sets are available upon request from the authors.
%
\bibitem{ref:hermessidis}  
  A.\ Hillenbrand,
  ``Measurement and Simulation of the Fragmentation Process at HERMES'',
  Ph.D.\ thesis, Erlangen Univ., Germany, September 2005;
  private communications.
%
\bibitem{ref:sia-data} 
  H.\ Aihara {\em et al.} [TP Collaboration],
  Phys. Rev. Lett. {\bf 61}, 1263 (1998); Phys. Lett. {\bf B184}, 299 (1987); 
  X.-Q.\ Lu, Ph.D.\ thesis, John Hopkins University, UMI-87-07273, 1986;
  K.\ Abe {\em et al.} [SLD Collaboration], Phys. Rev. {\bf D59}, 052001 (1999);
  D.\ Buskulic {\em et al.} [ALEPH Collaboration],  Z. Phys. {\bf C66}, 355 (1995);
  P.\ Abreu {\em et al.} [DELPHI Collaboration], Eur. Phys. J. {\bf C5}, 585 (1998);
  R.\ Akers {\em et al.} [OPAL Collaboration], Zeit. Phys. {\bf C63}, 181 (1994).
%
\bibitem{ref:phenixpion} 
  S.S.\ Adler {\em et al.} [PHENIX Collaboration] , 
  Phys. Rev. Lett. {\bf 91}, 241803 (2003); 
  A.~Adare {\it et al.}  [PHENIX Collaboration],
  Phys.\ Rev.\ D {\bf 76}, 051106 (2007).
%
\bibitem{ref:nlo} 
  B.~Jager, A.~Schafer, M.~Stratmann, and W.~Vogelsang,
  Phys.\ Rev.\ D {\bf 67}, 054005 (2003);
%
\end{thebibliography}
\end{document}